\documentclass[a4paper]{article}

\usepackage{INTERSPEECH2018}
\usepackage{url,multirow}
\ninept

\usepackage{color}

\usepackage{hyperref}

\makeatletter
\def\bstctlcite{\@ifnextchar[{\@bstctlcite}{\@bstctlcite[@auxout]}}
\def\@bstctlcite[#1]#2{\@bsphack
  \@for\@citeb:=#2\do{%
    \edef\@citeb{\expandafter\@firstofone\@citeb}%
    \if@filesw\immediate\write\csname #1\endcsname{\string\citation{\@citeb}}\fi}%
  \@esphack}
\makeatother 

\title{Scaling Speech Enhancement in Unseen Environments with Noise Embeddings}
\name{Gil Keren$^1$, Jing Han$^1$, Bj\"orn Schuller$^{1,2}$}
\address{\fontsize{11}{11}\selectfont
  $^1$ZD.B Chair of Embedded Intelligence for Health Care and Wellbeing, University of Augsburg, Germany\\
  $^2$GLAM -- Group on Language, Audio \& Music, Imperial College London, UK}
\email{cruvadom@gmail.com}

\begin{document}
\bstctlcite{IEEEexample:BSTcontrol}
  \maketitle

\begin{abstract}
We address the problem of speech enhancement generalisation to unseen environments by performing two manipulations. First, we embed an additional recording from the environment alone, and use this embedding to alter activations in the main enhancement subnetwork. Second, we scale the number of noise environments present at training time to 16,784 different environments. Experiment results show that both manipulations reduce word error rates of a pretrained speech recognition system and improve enhancement quality according to a number of performance measures. Specifically, our best model reduces the word error rate from 34.04\% on noisy speech to 15.46\% on the enhanced speech\footnote{Enhanced audio samples can be found in \url{https://speechenhancement.page.link/samples}}. 
\end{abstract}

\section{Introduction}
Speech processing in everyday life presents the challenge of obtaining good model performance across a large variety of environments. In fact, the variety of real-world environments is large enough to assume that a speech processing model would be required to perform well in unseen environments that do not closely resemble the ones present during its training stage. While many existing approaches to speech enhancement focus on a small number of environments or environments that are similar to each other \cite{DBLP:conf/interspeech/KumarF16,DBLP:conf/icassp/SeltzerYW13,Weninger15-SEW}, in contrast, in this work we aim at designing a speech enhancement model that performs well across a large variety of environments, many of which could be considerably different from the ones seen at training time. 

% What we do 
To this end, we explore two methods we expect to contribute to speech enhancement generalisation to unseen environments. First, inspired by one-shot learning models \cite{keren2018weakly}, we do not think of the noise environments used in this work as distinct unrelated categories, but rather as samples from a large space that contains all noise environments. In this setting, good speech enhancement in an unseen noise environment would amount to generalising to an unseen point in the noise environments space. Therefore, to facilitate good generalisation in this space, we scale up the training set size to include noises from 16,784 different environments, mixed with 360 hours of clean speech and different Signal-to-Noise Ratios (SNRs). As the number of environments seen during training is large, we expect those environments to share some properties with the unseen test environments, that may assist in the enhancement process.
 
Moreover, we hypothesise that providing the network with an additional recording of the same environment may assist the network in identifying which frequency components need to be removed and which need to be enhanced. Specifically, we condition our model on an additional recording that contains no speech, from the same environment as the noisy speech segment. A dedicated subnetwork processes this additional recording to create a \emph{noise embedding}, that is in turn injected to all layers of the main enhancement subnetwork. This is a plausible scenario, as enhancement devices may record an environment noise sample just before recording the noisy speech.

In experiments, we show that both manipulations result in better speech enhancement compared to baseline methods, as measured by the Word Error Rate (WER) of a pretrained speech recognition system, as well as a number of objective evaluation metrics. Specifically, while WER on noisy speech with SNR of 0-25 dB was 34.04\%, using our enhancement model with as little as 200 training noise environments and no noise embeddings reduced WER for unseen noise environments to 21.51\%. Scaling up the  the number of training noise environments to 16,784 managed to reduce WER to 16.78\%. Finally, using the noise embeddings computed from additional environment recordings reduced WER to 15.46\%. 

\section{Data generation} \label{sec:data}

% Training set is massive
As motivated above, we aim to improve audio enhancement generalisation to unseen environments by training an enhancement model with a large number of environment noises. To this end, we mix clean speech utterances from the Librispeech corpus \cite{panayotov2015librispeech} with different noise recordings from the recently published Audio Set \cite{gemmeke2017audio}. 

The Audio Set corpus contains 2,100,000 audio segments of 10 seconds extracted from YouTube videos, manually annotated in a multi-label manner for a hierarchical ontology of 632 audio categories, covering a wide range of human and animal sounds, musical instruments and genres, and common everyday environmental sounds. Audio Set contains recordings of a large number noises with great diversity over noise sources and recording conditions, therefore is challenging and may be a good approximation of the set of noises coming from natural everyday environments.
We use noises from all categories in Audio Set's ontology, excluding noises that are labelled as `Human sounds' or `Music' (as those may contain speech). The clips were selected randomly, in a manner that balances between the amount of noises from each of the top categories in the ontology. We found that a small portion of the noises that were not labelled as containing human speech did contain utterances of human speech anyway.

\begin{figure*}[t]
  \centering
  \includegraphics[width=\linewidth]{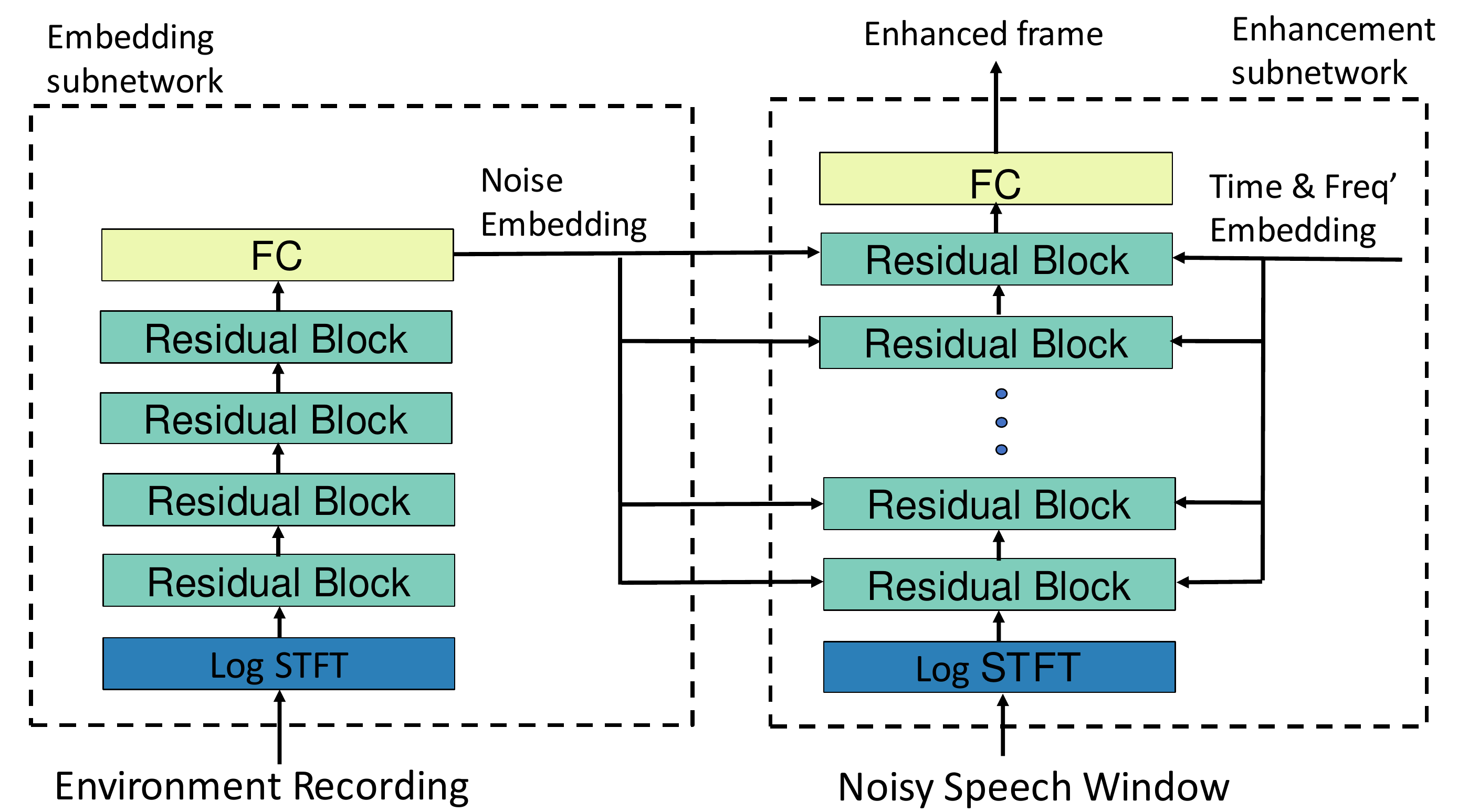}
  \caption{The enhancement model architecture. The embedding subnetwork processes an environment recording through a sequence of residual blocks to produce a fixed length vector that is the noise embedding. In the main enhancement subnetwork, the noisy speech recording is processed through another sequence of residual blocks, each conditioned on the noise embedding and additional positional embeddings, to emit one enhanced speech frame.}
  \label{fig:model}
\end{figure*}

To make sure that at test time we enhance audio from unseen environments, each noise segment is assigned to either the training, validation or test set. In total, in our training set we use 16,784 different recordings of different environments/noises, 656 in the validation set and 740 in the test set.
The clean speech utterances we mix with environment noises were taken from the Librispeech corpus. The corpus is freely available for downloading, and includes a train, validation and test splits. The Librispeech training set is comprised of 360 hours of audio from 921 speakers, where the validation and test sets contain 5.4 hours of speech and 40 speakers each. For the enhancement task training, validation and test sets, we mix clean utterances from the corresponding set in the Librispeech corpus.

Clean speech utterances are mixed with the environment recordings using SNRs of 0, 5, 10, 15, 20 and 25 dB. For the training set, during training a random speech utterance is mixed with a random noise recording with a random SNR, to obtain the largest effective training set size. For the validation and test sets, one environment noise recording and SNR are fixed for every utterance.

\section{Enhancement model} \label{sec:model}

We design a neural network enhancement model, that can be conditioned on environment noise samples and has a large capacity that corresponds to the large training set. We represent all audio segments as their log magnitude spectrum, which is obtained by taking the log absolute value of the Short-Time Fourier Transform (STFT) over 25\,ms frames shifted by 10\,ms, where $10^{-7}$ is added before computing the log function to prevent the model from trying to fit unimportant differences in magnitude. All audio recordings we use have a sample rate of 16kHz, which results in 201 coefficients for each frame. 

Our enhancement model processes two inputs: $n$ frames of \textit{noisy speech segment} (speech mixed with noise, as described in Section \ref{sec:data}) and $r$ frames that are the \textit{noise segment} (in our experiments we found the best values are $n=200$ and $r=35$). It is important to note that in all cases the noise segment and the noise that was mixed into the noisy speech segment are from the same recording, but from different parts of this recording. This better simulates a real-world situation, where an enhancement device may record the environment sound alone, just before recording the noisy speech. The network processes the two inputs and then outputs the \textit{enhanced frame}. The enhanced frame is the network approximation for the central frame of the clean speech recording (before mixing it with noise), which we refer to as the \textit{clean frame}. A diagram of our enhancement model architecture is found in Figure \ref{fig:model}.

\vspace{-0.1cm}
\subsection{Embedding subnetwork}
\vspace{-0.05cm}
We first process the noise segment through an embedding subnetwork to create the \text{noise embedding}. The embedding subnetwork is comprised of a sequence of 4 residual blocks \cite{he2016deep}, each comprised of two 2D-convolutional layers (we treat the time and frequency as spatial axes). The convolutional layers of the first two residual blocks use a kernel size of $8 \times 4$ (time steps over frequency components) with a stride of $3 \times 2$, where the convolutional layers of the last two residual blocks use a $4 \times 4$ kernel with a $1 \times 1$ and $1 \times 2$ stride respectively. The number of feature maps for the convolutional layers in each of the four residual block are 64, 128, 256, 512 respectively. Inside each residual block, batch normalisation \cite{DBLP:conf/icml/IoffeS15} is applied on the output of the first convolutional layer, followed by a rectified linear activation function. Then, the second convolutional layer is applied and its output is added to the block's input. Batch normalisation is then applied again, followed by another rectified linear activation function to return the block's output. After applying the four residual blocks, we average each feature map across all locations, to obtain a single 512-dimensional vector that is the noise embedding.

\vspace{-0.1cm}
\subsection{Enhancement subnetwork}

The enhancement subnetwork then processes the noisy speech segment and the noise embedding, to output the enhanced frame. This subnetwork is comprised of a sequence of 8 residual blocks, processing the noisy speech segment. In those residual blocks, for each convolutional layer, the noise embedding is linearly projected to a dimension equal to the number of feature maps in this convolution layer. This projected noise embedding is then added to each location in the output map of the convolution layer. By doing so, we inject the noise embedding along all parts of the deep network, allowing each part to emit different outputs based on the current environment noise that should be removed from the audio. 

Additionally, to allow the network to process differently the different frequency components and time steps, we embed the time steps and frequency components (location embeddings) indices and add these embeddings to the appropriate locations in the output map of each convolutional layer. The location embeddings are emitted using a small neural network. The input of this network is a single integer that represents the location in the appropriate axis (time / frequency), which is processed through two 50-dimensional fully-connected layers with rectified linear activation function and batch normalisation. Except the described above, the structure of each residual block in this subnetwork is identical to those in the embedding subnetwork.

The convolutional layers in the first 4 residual blocks of the enhancement subnetwork use a $4 \times 4$ kernel, where a $3 \times 3$ kernel is used in the last 4 residual blocks. A $2 \times 2$ stride is applied in residual blocks number 3, 5 and 7. The first two residual blocks use 64 feature maps for each convolutional layer, followed by 128, 256 and 512 feature maps for the next groups of 2 residual blocks.

For the final part of the enhancement subnetwork, we flatten the output of the last residual block and feed it into a fully-connected layer with 201 output values. The output of this layer is treated as an \textit{enhancement mask}, which is added to the central frame of the noisy speech segment to yield the enhanced frame. During training, network parameters are optimised to minimise the mean squared error between the enhanced frame and the clean frame. In our experiments we used Stochastic Gradient Descent (SGD), where $0.1$ was found to be the best learning rate. 

\section{Experiments}

We conduct a series of experiments to study the effect of noise embeddings and the number of different environment noises available in training time on speech enhancement performance. 

\begin{table}[!h]
\centering
\caption{Comparison of test set evaluation metrics for all enhancement models and the noisy speech. `No Emb x' stands for the model with no noise embedding where x is the number of environment noises available at training time. Best results in bold. Performance in all evaluation metrics improves with the increase in the number of training environments recordings, and using noise embedding further improves performance.  }
\label{tab:results}
\begin{tabular}{|l|r|r|r|r|}
\toprule
Method & WER & PESQ & SegSNR & LSD \\
\midrule
Clean Speech 			& 4.21	& ---  & ---   & ---  \\
\midrule
Noisy Speech			& 34.04	& 2.59 & 7.02  & 0.94 \\
\midrule
Log-MMSE 				& 35.38	& 2.66 & 7.12  & 0.91 \\
\midrule
Noise Aware				& 25.30	& 2.96 & 11.01 & 0.54 \\
\midrule
Ours - No Emb' 200		& 21.51	& 3.12 & 10.03 & 0.53 \\
\midrule
Ours - No Emb' 1000		& 20.54	& 3.13 & 10.00 & 0.52 \\
\midrule
Ours - No Emb' All		& 16.78	& 3.25 & 11.71 & 0.48 \\
\midrule
Ours - With Emb' 	& \textbf{15.46}	& \textbf{3.30} & \textbf{12.99} & \textbf{0.45} \\
\bottomrule
\end{tabular}
\end{table}

\begin{table*}[!h]
\centering
\caption{Test set word error rates using the pretrained speech recognition system, the different enhancement methods and SNRs ranging from 0 to 25 dB. For all SNRs, best performance is obtained using the largest number of training set noises and noise embeddings.}
\label{tab:wer}
\begin{tabular}{|l|r|r|r|r|r|r|}
\toprule
Method & $0$ dB & $5$ dB & $10$ dB & $15$ dB & $20$ dB & $25$ dB \\
\midrule
Noisy Speech			& 75.09 & 57.22 & 37.54 & 18.98 & 8.83 & 5.19 \\
\midrule
Log-MMSE 				& 77.25 & 59.90 & 37.49 & 19.93 & 10.01 & 6.21 \\
\midrule
Noise Aware				& 61.41 & 40.15 & 24.84 & 12.38 & 6.72 & 4.93 \\
\midrule
Ours - No Embeddings 200		& 54.24 & 31.95 & 20.36 & 10.34 & 6.13 & 4.75 \\
\midrule
Ours - No Embeddings 1000		& 52.97 & 29.95 & 18.77 & 9.47 & 5.98 & 4.80 \\
\midrule
Ours - No Embeddings All		& 43.85 & 22.92 & 15.00 & 7.64 & 5.50 & 4.64 \\
\midrule
Ours - With Embeddings 		& \textbf{41.76} & \textbf{21.34} & \textbf{12.35} & \textbf{6.88} & \textbf{5.18} & \textbf{4.05} \\
\bottomrule
\end{tabular}
\end{table*}

% Converting to wav and evaluation metrics
Inverse Short-Time Fourier Transform (ISTFT) was applied to the enhanced magnitude spectrogram, together with the phase of the noisy speech, to reconstruct a waveform from each enhanced log magnitude spectrum. To measure the success of each of our speech enhancement models, we used the following evaluation metrics. The first evaluation metric is the WER obtained by feeding the enhanced audio to a pretrained `Deep Speech' speech recognition system \cite{hannun2014deep}, where a model that was trained on the Librispeech dataset can be found in \url{https://github.com/mozilla/DeepSpeech}. In addition, we computed three other well established metrics for the assessment of speech audio quality: the Perceptual Evaluation of Speech Quality (PESQ) \cite{DBLP:conf/icassp/RixBHH01}, that is an industry standard for objective voice quality testing, Segmental Signal-to-Noise Ratio (SegSNR) \cite{DBLP:conf/interspeech/HansenP98} and Log-Spectral Distortion (LSD) \cite{LSD}, that evaluate the clean speech features reconstruction in different manners. 
 
% Different models we evaluate
We compared the enhancement performance of several models. First, we evaluated our hypothesis that speech enhancement performance in unseen environments is closely related to the number of different environment noises appearing during training time. To this end, we compared three identical enhancement networks, trained with utterances that are mixed with 200, 1000, and our full training set of 16,784 environment noises respectively. These enhancement networks do not use the noise embedding, which is otherwise added to the output of the convolutional layers.
Second, to study the effect of the noise embedding, we trained another identical enhancement network with the full training set of environment noises, that does use the noise embedding as described in Section \ref{sec:model}. 

Finally, we compared our proposed network with two other baselines, Log-MMSE and noise-aware training. Log-MMSE \cite{DBLP:journals/tsp/EphraimM85} is a more traditional, non-neural network enhancement method, while in noise-aware training \cite{DBLP:conf/icassp/SeltzerYW13} an estimation of the noise is given to a fully-connected enhancement network as additional features. In our case, we average frames of the noise segment along the time axis to create the noise estimation used in noise aware training, in a manner inspired from \cite{DBLP:conf/icassp/SeltzerYW13}. We optimise the number of layers (3-7), hidden units (500-2000) and regularisation (batch normalisation) in the fully-connected network used in noise-aware training. In all cases, model parameters were chosen using the validation set, for each method separately. The training, validation and test sets all contain different environment noises, to assess the model's performance in unseen environments.

% Results and interpretations
Test set results for all enhancement methods and the noisy speech can be found in Table \ref{tab:results}, averaged across all SNRs. Our test set contains noisy speech from noise environments, speakers and utterances that did not appear in training time. First, we observe that the pretrained speech recognition system obtained a 4.21\% WER on the clean speech, while obtaining 34.04\% on the noisy speech before applying any enhancement method. The Log-MMSE baseline did not manage to considerably improve over the noisy speech, and for the WER evaluation metric even caused a degradation in performance. The other baseline method, noise-aware training, did manage to considerably improve all evaluation metrics over the noisy speech, with a 25.30\% WER. 

Further, we compared our proposed method to the baseline methods. Using our enhancement network with only 200 training noise environments and no noise embedding managed to further reduce WER to 21.51\%, and improve all other evaluation metrics in a similar manner. This finding indicates that the structure of the enhancement network is important, and a deep, residual network is preferable over a fully-connected network. Next, we observe that further increasing the number of training noise environments causes a substantial improvement in all evaluation metrics, with a WER of 16.78\% when using all 16,784 training noise environments. Last, using the embedding subnetwork and feeding the noise embedding to the enhancement model as described in Section \ref{sec:model} reduces WER to 15.46\% and improves all other evaluation metrics as well. The 15.46\% WER that is obtained by our best model corresponds to a relative reduction of 54.58\% in WER, compared to the original noisy speech. 

A decomposition of WERs for the different enhancement methods and SNRs can be found in Table \ref{tab:wer}. The results in the table show that the same conclusions we draw from Table \ref{tab:results} also hold for each SNR independently. Specifically, we observe the advantages of our methods that use a deep residual network over the baseline methods, and the improvement in WER due to a largest number of training noises and the usage of noise embeddings. Moreover, when using our best model to enhance noisy speech in the 25 dB condition, we surprisingly found that the WER obtained on the enhanced speech (4.05\%) is better even compared to the WER obtained on the clean speech recordings (4.21\%). This finding may indicate that the enhancement model captures the dynamics of speech up to a level of denoising seemingly negligible background noises that exist in the clean recordings, and further enhances speech quality in those recordings.

\section{Conclusions}
We investigated speech enhancement in unseen environments using two main manipulations. First, we view speech enhancement in unseen environments as the problem of generalising to unseen points in the space of noise environments. Motivated by this, we scale the number of training noise environments to 16,784. Second, we supply the enhancement model with additional information about the environment, in the form of an additional recording from the noise environment. This additional recording is processed to create a noise embedding vector that is fed as an additional input to different layers of the main enhancement subnetwork. In experiments, we observed that both of our manipulations managed to considerably improve the quality of the speech enhancement model, as measured by a variety of evaluation metrics. For example, our best model manages to reduce the WER of a pretrained speech recognition system from 34.04\% on noisy speech to 15.46\% on the enhanced speech. 

In future work we plan on further exploring the method of additional context embedding, e.g., embedding speaker recordings for source separation and embedding noises for selective denoising, as well as improving the current enhancement model using alternative training mechanisms \cite{keren2018fast,keren2017tunable}. In addition, we plan on exploring the resulting embedding spaces for semantic and acoustic interpretations. 

\section{Acknowledgements}
This work was supported by Huawei Technologies Co. Ltd. and the European
Union's Seventh Framework Programme through the ERC Starting Grant No. 338164 (ERC StG iHEARu).

\nocite{*}
  % Bibliography
\eightpt
\bibliographystyle{IEEEtran}
\bibliography{mybib}

\begin{thebibliography}{10}
\providecommand{\url}[1]{#1}
\csname url@samestyle\endcsname
\providecommand{\newblock}{\relax}
\providecommand{\bibinfo}[2]{#2}
\providecommand{\BIBentrySTDinterwordspacing}{\spaceskip=0pt\relax}
\providecommand{\BIBentryALTinterwordstretchfactor}{4}
\providecommand{\BIBentryALTinterwordspacing}{\spaceskip=\fontdimen2\font plus
\BIBentryALTinterwordstretchfactor\fontdimen3\font minus
  \fontdimen4\font\relax}
\providecommand{\BIBforeignlanguage}[2]{{%
\expandafter\ifx\csname l@#1\endcsname\relax
\typeout{** WARNING: IEEEtran.bst: No hyphenation pattern has been}%
\typeout{** loaded for the language `#1'. Using the pattern for}%
\typeout{** the default language instead.}%
\else
\language=\csname l@#1\endcsname
\fi
#2}}
\providecommand{\BIBdecl}{\relax}
\BIBdecl

\bibitem{DBLP:conf/interspeech/KumarF16}
A.~Kumar and D.~A.~F. Florencio, ``Speech enhancement in multiple-noise
  conditions using deep neural networks,'' in \emph{Proc. of {INTERSPEECH}},
  2016, pp. 3738--3742.

\bibitem{DBLP:conf/icassp/SeltzerYW13}
M.~L. Seltzer, D.~Yu, and Y.~Wang, ``An investigation of deep neural networks
  for noise robust speech recognition,'' in \emph{Proc. of {ICASSP}}, 2013, pp.
  7398--7402.

\bibitem{Weninger15-SEW}
F.~Weninger, H.~Erdogan, S.~Watanabe, E.~Vincent, J.~{Le Roux}, J.~R. Hershey,
  and B.~Schuller, ``{Speech Enhancement with LSTM Recurrent Neural Networks
  and its Application to Noise-Robust ASR},'' in \emph{{Proceedings of the
  International Conference on Latent Variable Analysis and Signal Separation}},
  vol. 9237, Liberec, Czech Republic, 2015, pp. 91--99.

\bibitem{keren2018weakly}
G.~Keren, M.~Schmitt, T.~Kehrenberg, and B.~Schuller, ``Weakly supervised
  one-shot detection with attention {S}iamese networks,'' \emph{arXiv preprint
  arXiv:1801.03329}, 2018.

\bibitem{panayotov2015librispeech}
V.~Panayotov, G.~Chen, D.~Povey, and S.~Khudanpur, ``Librispeech: {A}n {ASR}
  corpus based on public domain audio books,'' in \emph{Proc. of {ICASSP}},
  2015, pp. 5206--5210.

\bibitem{gemmeke2017audio}
J.~F. Gemmeke, D.~P. Ellis, D.~Freedman, A.~Jansen, W.~Lawrence, R.~C. Moore,
  M.~Plakal, and M.~Ritter, ``Audio {S}et: {A}n ontology and human-labeled
  dataset for audio events,'' in \emph{Proc. of {ICASSP}}, 2017, pp. 776--780.

\bibitem{he2016deep}
K.~He, X.~Zhang, S.~Ren, and J.~Sun, ``Deep residual learning for image
  recognition,'' in \emph{Proc. of {CVPR}}, 2016, pp. 770--778.

\bibitem{DBLP:conf/icml/IoffeS15}
S.~Ioffe and C.~Szegedy, ``Batch normalization: Accelerating deep network
  training by reducing internal covariate shift,'' in \emph{Proceedings of
  {ICML}}, Lille, France, 2015, pp. 448--456.

\bibitem{hannun2014deep}
A.~Hannun, C.~Case, J.~Casper, B.~Catanzaro, G.~Diamos, E.~Elsen, R.~Prenger,
  S.~Satheesh, S.~Sengupta, A.~Coates \emph{et~al.}, ``Deep speech: Scaling up
  end-to-end speech recognition,'' \emph{arXiv preprint arXiv:1412.5567}, 2014.

\bibitem{DBLP:conf/icassp/RixBHH01}
A.~W. Rix, J.~G. Beerends, M.~P. Hollier, and A.~P. Hekstra, ``Perceptual
  evaluation of speech quality {PESQ}-a new method for speech quality
  assessment of telephone networks and codecs,'' in \emph{Proc. of {ICASSP}},
  2001, pp. 749--752.

\bibitem{DBLP:conf/interspeech/HansenP98}
J.~H.~L. Hansen and B.~L. Pellom, ``An effective quality evaluation protocol
  for speech enhancement algorithms,'' in \emph{Proc. of {ICSLP}}, 1998.

\bibitem{LSD}
A.~Gray and J.~Markel, ``Distance measures for speech processing,''
  \emph{{IEEE} Trans. Acoustics, Speech, and Signal Processing}, vol.~24,
  no.~5, pp. 380--391, 1976.

\bibitem{DBLP:journals/tsp/EphraimM85}
Y.~Ephraim and D.~Malah, ``Speech enhancement using a minimum mean-square error
  log-spectral amplitude estimator,'' \emph{{IEEE} Trans. Acoustics, Speech,
  and Signal Processing}, vol.~33, no.~2, pp. 443--445, 1985.

\bibitem{keren2018fast}
G.~Keren, S.~Sabato, and B.~Schuller, ``Fast single-class classification and
  the principle of logit separation,'' in \emph{Proceedings of {ICDM}},
  Singapore, 2018, to appear.

\bibitem{keren2017tunable}
G.~Keren, S.~Sabato, and B.~Schuller, ``Tunable sensitivity to large errors in
  neural network training,'' in \emph{Proceedings of {AAAI}}, San Francisco,
  CA, 2017, pp. 2087--2093.

\end{thebibliography}

\end{document}